\documentclass[aps,prl,twocolumn,showpacs,groupedaddress,floatfix]{revtex4}  
\usepackage{graphicx}  
\usepackage{dcolumn}   
\usepackage{bm}        
\usepackage{amssymb}   
\usepackage{psfrag}
\usepackage{multirow}
\usepackage{dcolumn}

\newcommand {\whatsthis}         {Letter} 

\newcommand {\gef}	{\ensuremath{\geqslant 5}}
\newcommand {\gefour}	{\ensuremath{\geqslant 4}}
\newcommand {\aobs}	{\ensuremath{A_{\text {fb}}^{\text{obs}}}}

\newcommand {\asym}	{\ensuremath{A_{\text {fb}}}}
\newcommand {\apred}	{\ensuremath{A_{\text {fb}}^{\text {pred}}}}
\newcommand {\nomA}	{\ensuremath{12}}
\newcommand {\fitA}	{\ensuremath{\pm \fitAnet}}
\newcommand {\fitAnet}	{\ensuremath{8}} 
\newcommand {\nomAF}	{\ensuremath{19}}
\newcommand {\fitAF}	{\ensuremath{\pm \fitAFnet}}
\newcommand {\fitAFnet}	{\ensuremath{9}} 
\newcommand {\nomAM}	{\ensuremath{-16}}
\newcommand {\fitAM}	{\ensuremath{^{+15}_{-17}}} 
\newcommand {\systA}	{\ensuremath{\pm 1}}
\newcommand {\systAF}	{\ensuremath{\pm 2}}
\newcommand {\systAM}	{\ensuremath{\pm 3}}

\newcommand {\dil}         {\ensuremath{{\cal D}}}

\newcommand {\WArecoSc}	{\ensuremath{4.4\pm1.6}}

\newcommand {\pbar}	{\ensuremath{\bar p}}
\newcommand {\ppbar}	{\ensuremath{p\pbar}}

\newcommand {\tbar}     {\ensuremath{\bar t}}

\newcommand {\ttbar}    {\ensuremath{t\tbar}}
\newcommand {\ttbarx}   {\ensuremath{\ttbar+X}}
\newcommand {\W}        {\ensuremath{W}}
\newcommand {\Wbos}     {\ensuremath{W}~boson}
\newcommand {\Wdbos}     {\ensuremath{W}~boson} 

\newcommand {\Z}        {\ensuremath{Z}}
\newcommand {\Zp}       {\ensuremath{Z'}}

\newcommand {\wpj}      {\W+jets}

\newcommand {\zpj}      {\Z+jets}
\newcommand {\lpj}      {lepton+jets}

\newcommand {\GeV}        {\ensuremath{\,\text {GeV}}}
\newcommand {\TeV}        {\ensuremath{\,\text {TeV}}}

\newcommand {\fbo}        {\ensuremath{\,\text {fb}^{-1}}}

\newcommand {\pt}         {\ensuremath{p_T}}
\newcommand {\njet}       {\ensuremath{N_{\text {jet}}}}

\newcommand {\dy}        {\ensuremath{\Delta y}}
\newcommand {\dyrec}        {\ensuremath{\Delta y_{\text {reco}}}}
\newcommand {\ady}        {\ensuremath{|\dy|}}
\newcommand {\dygen}        {\ensuremath{\Delta y}}
\newcommand {\adYgen}        {\ensuremath{\left|\dygen\right|}}
\newcommand{\emiss}      {/\!\!\!\!E} 
\newcommand{\met}        {\ensuremath{\emiss_T}}
\newcommand{\ld}         {\ensuremath{{\cal {L}}}}

\newcommand {\NFwd}       {\ensuremath{N_{\text{f}}}}
\newcommand {\NBwd}       {\ensuremath{N_{\text{b}}}}

\newcommand {\DZ}         {D0} 
\newcommand {\CDF}        {CDF}
\newcommand {\stat} 	{\ensuremath{\left(\mathrm{stat.} \right)}}
\newcommand {\syst} 	{\ensuremath{\left(\mathrm{syst.} \right)}}
\newcommand {\acce} 	{\ensuremath{\left(\mathrm{accept.}\right)}}
\newcommand {\dilu} 	{\ensuremath{\left(\mathrm{dilution}\right)}}

\newcommand{\fullWid}	{0.98}
\newcommand{\halfWid}	{0.48}

 \newcommand {\pythia}   {{\sc pythia}}
 \newcommand {\geant}    {{\sc geant}}
 \newcommand {\alpgen}   {{\sc alpgen}}
 \newcommand {\mcatnlo}  {{\sc mc@nlo}}
 \newcommand {\pxcone}   {{\sc pxcone}}

 \newcommand {\eom}      {electron or muon}
 \newcommand {\etal}     {{\it et al.}}

\psfrag {LD2} {\ld}

\begin{document}

\hspace{5.2in} \mbox{FERMILAB-PUB-07-645-E}

\title{First measurement of the forward-backward charge asymmetry in top quark pair production}

%
\author{V.M.~Abazov$^{36}$}
\author{B.~Abbott$^{76}$}
\author{M.~Abolins$^{66}$}
\author{B.S.~Acharya$^{29}$}
\author{M.~Adams$^{52}$}
\author{T.~Adams$^{50}$}
\author{E.~Aguilo$^{6}$}
\author{S.H.~Ahn$^{31}$}
\author{M.~Ahsan$^{60}$}
\author{G.D.~Alexeev$^{36}$}
% ...

\affiliation{\vspace{0.1 in}(The D\O\ Collaboration)\vspace{0.1 in}}
% ...
\affiliation{$^{6}$University of Alberta, Edmonton, Alberta, Canada,
                Simon Fraser University, Burnaby, British Columbia, Canada,
                York University, Toronto, Ontario, Canada, and
                McGill University, Montreal, Quebec, Canada}
% ...
\affiliation{$^{29}$Tata Institute of Fundamental Research, Mumbai, India}
% ...
\affiliation{$^{31}$Korea Detector Laboratory, Korea University, Seoul, Korea}
% ...
\affiliation{$^{36}$Joint Institute for Nuclear Research, Dubna, Russia}
% ...
\affiliation{$^{50}$Florida State University, Tallahassee, Florida 32306, USA}
% ...
\affiliation{$^{52}$University of Illinois at Chicago,
                Chicago, Illinois 60607, USA}
% ...
\affiliation{$^{60}$Kansas State University, Manhattan, Kansas 66506, USA}
% ...
\affiliation{$^{66}$Michigan State University,
                East Lansing, Michigan 48824, USA}
% ...
\affiliation{$^{76}$University of Oklahoma, Norman, Oklahoma 73019, USA}
% ...

\date{December 5, 2007}
\begin{abstract}
We present the first measurement of 
the integrated forward-backward charge asymmetry in top-antitop 
quark pair (\ttbar) production in proton-antiproton (\ppbar) collisions
in the \lpj\ final state.
Using a $b$-jet tagging algorithm and kinematic
reconstruction assuming \ttbarx\ production and decay,
a sample of $0.9\fbo$ of data, collected 
by the \DZ\ experiment
at the Fermilab Tevatron Collider,
is used to measure the asymmetry for different jet multiplicities.
The result is also used to set upper limits on 
\ttbarx\ production via a \Zp\ resonance.
\end{abstract}

\maketitle 


At lowest order in quantum chromodynamics (QCD), 
the standard model (SM) predicts that 
the kinematic distributions in $\ppbar\to\ttbarx$
 production are charge symmetric.
But this symmetry is accidental, as the initial
\ppbar\ state is not an eigenstate of charge conjugation.
Next-to-leading order (NLO) calculations predict 
forward-backward asymmetries of (5--10)\%~\cite{ref:asymwhy,ref:asymtev},
but recent
next-to-next-to-leading order (NNLO) calculations 
predict significant corrections for $\ttbar$
production in association with a jet~\cite{ref:asymNNLO}.
The asymmetry arises mainly from interference
between contributions symmetric and antisymmetric 
under the exchange $t \leftrightarrow \tbar$~\cite{ref:asymwhy},
and depends on the region of phase space being
probed and, in particular, 
on the production of an additional jet~\cite{ref:asymtev}.
The small asymmetries expected in the SM
make this a sensitive probe for new 
physics~\cite{ref:axigluon}.

A charge asymmetry in $\ppbar\to\ttbarx$
can be observed as a forward-backward production asymmetry.
The signed difference between the
rapidities~\cite{ref:eta} of the $t$ and \tbar,
$\dy \equiv y_t - y_{\tbar}$, reflects the asymmetry in \ttbar\ production.
We define the integrated charge asymmetry as
$
\asym = \left(\NFwd-\NBwd\right) / \left(\NFwd+\NBwd\right),
$
where \NFwd\ (\NBwd) is the number of events with 
a positive (negative) \dy.

This \whatsthis\ describes the first measurement of 
\asym\ in $\ppbar\to\ttbarx$ production.
The $0.9\fbo$ data sample used
was collected at $\sqrt{s} = 1.96\TeV$ 
with the \DZ\ detector~\cite{ref:d0det}, using
triggers that required a jet and an electron or muon.
In the \lpj\ final state of the \ttbar\ system,
one of the two \Wbos s from the \ttbar\ pair 
decays into hadronic jets and the other into leptons,
yielding a signature of two $b$-jets, 
two light-flavor jets, an isolated lepton, 
and missing transverse energy (\met).
This decay mode is well suited for this measurement,
as it
combines a large branching fraction 
($\sim\!34\%$) with high signal purity, 
the latter a consequence of requiring an isolated \eom\ with large
transverse momentum (\pt).
The main background is from \wpj\ and multijet production.
This channel allows accurate reconstruction of the
$t$ and \tbar\ directions in the collision rest frame,
and the charge of the \eom\ 
distinguishes between the $t$ and \tbar\ quarks.

\begin{figure}
\vspace{-0.45cm}
\includegraphics[width=\fullWid\linewidth]{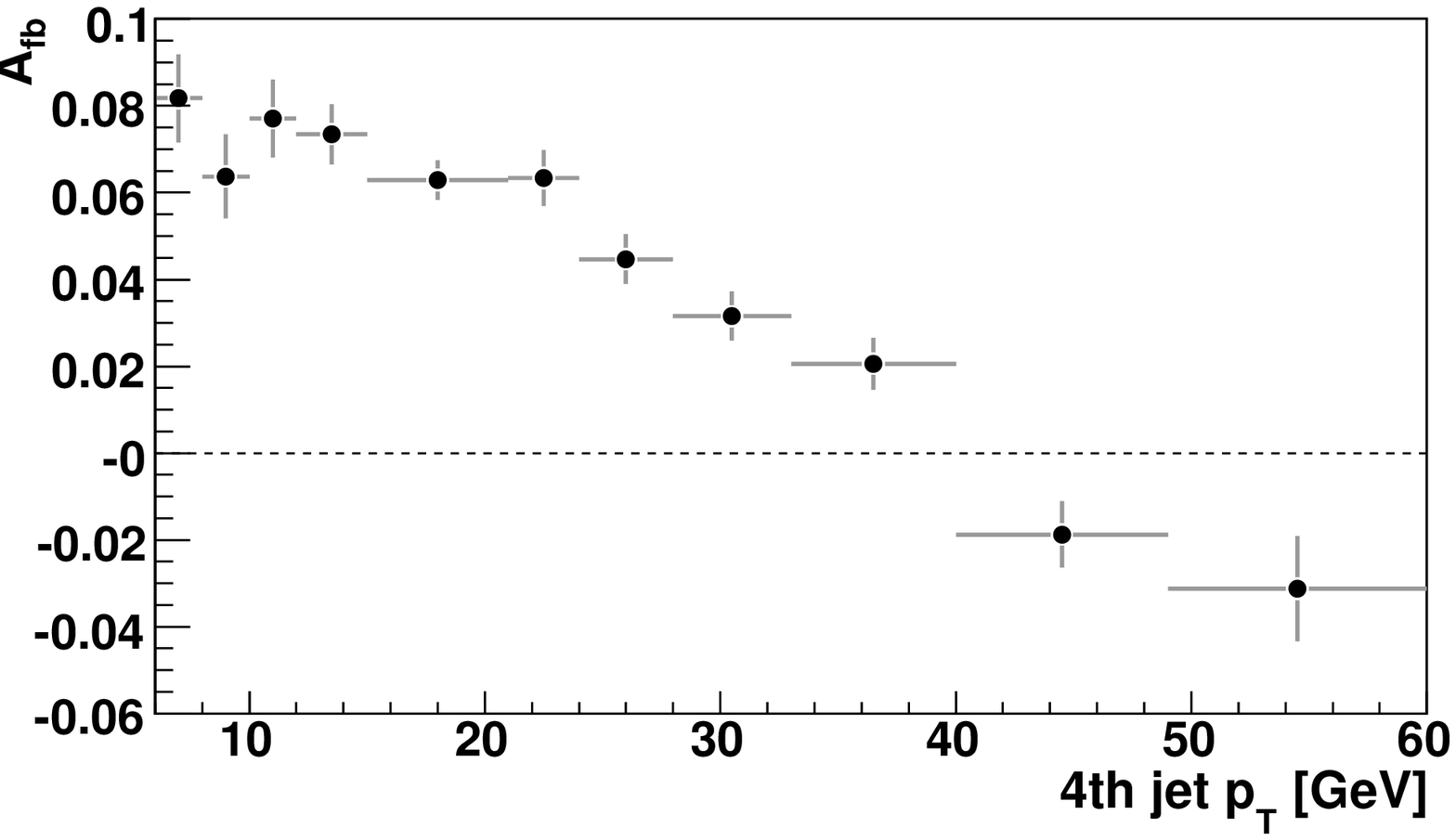}
\vspace{-4.6cm}
\begin{center}
{\bf \small MC@NLO} 
\end{center}
\vspace{3.1cm}
\caption{
Forward-backward \ttbar\ charge asymmetry predicted by
\mcatnlo\ as a function of the fourth-highest particle jet \pt.
}
\label{fig:diffAsym}
\end{figure}

The dependence of \asym\ on the region of phase space,
as calculated by the \mcatnlo\ event generator~\cite{ref:mcatnlo}, 
is demonstrated in Fig.\ \ref{fig:diffAsym}.
The large dependence on the fourth-highest jet \pt\ is
not available in the calculations of Refs.\ \cite{ref:asymwhy,ref:asymtev,ref:asymNNLO},
as these do not consider decays of the top quarks, 
and include only acceptance for jets from additional radiation.

We conclude that acceptance can strongly affect the asymmetry.
To facilitate comparison with theory, the analysis is therefore designed to have
an acceptance which can be described simply. 
Event selection is limited to either:
(i) selections on directions and momenta
that can be described at the particle level (which refers to
produced particles before they start interacting
with material in the detector) or
(ii) criteria with high signal efficiency, so that
their impact on the region of acceptance is negligible.
In addition, the observable quantity and the fitting procedure 
are chosen to ensure that all events have
the same weight in determining the asymmetry.

The measurement is not corrected for acceptance and
reconstruction effects, but a prescription provides
the acceptance at the particle level. 
Reconstruction effects 
are also accommodated at the particle level by
defining the asymmetry as a function of the generated \ady:
\begin{equation}
\asym\left(\ady\right) = \frac {g(\ady)-g(-\ady)}{g(\ady)+g(-\ady)},
\end{equation}
where $g$ is the probability density for \dy\ within the acceptance.
This asymmetry can be folded with the ``geometric dilution,'' 
\dil, which is described later:
\begin{equation}
\apred = \int_{0}^{\infty} \asym\left(\dy\right) 
\dil\left(\dy\right) \left[ g\left(\dy\right) + g\left(-\dy\right) \right]\,d\dy.
\label{eq:applyDil}
\end{equation}
This procedure yields the predictions in Table\ \ref{tab:pred}.
The values are smaller than those of Ref.\ \cite{ref:asymwhy,ref:asymtev},
because of the inclusion of jet acceptance and dilution.

\begin{table}
\caption{\label{tab:pred}
Predictions based on \mcatnlo.
}
\begin{tabular*}{\linewidth}{l@{\extracolsep{\fill}}r@{\,$\pm$\,\extracolsep{0pt}}l@{\stat\,$\pm$}l@{\acce\,$\pm$\,}l@{\dilu}}
\hline
\hline	
\njet & \multicolumn{4}{c}{\apred\ (in \%)} \\
\hline
$\geqslant 4$ &   0.8  & 0.2 & 1.0 & 0.0\\
$4$           &   2.3  & 0.2 & 1.0 & 0.1\\
\gef          & $-4.9$ & 0.4 & 1.0 & 0.2\\
\hline
\hline
\end{tabular*}
\end{table}

We select events with at least four jets reconstructed
using a cone algorithm~\cite{ref:d0jets} with 
an angular radius 
${\mathcal R} = 0.5$ (in rapidity and azimuthal angle).
All jets must have $\pt>20\GeV$ and pseudorapidity
(relative to the reconstructed primary vertex) $|\eta|<2.5$.
The leading jet must have $\pt>35\GeV$.
Events are required to have $\met>15\GeV$ and exactly
one isolated electron with $\pt>15\GeV$ and $|\eta|<1.1$
or one isolated muon with $\pt>18\GeV$ and $|\eta|<2.0$.
More details on lepton identification and trigger requirements
are given in Ref.\ \cite{ref:matrixmethod}. 
Events in which the lepton momentum is mismeasured are suppressed
by requiring that the direction of the \met\ not be along
or opposite the azimuth of the lepton. 
To enhance the signal, at least one of the jets is required to be
identified as originating from a long-lived $b$ hadron by
a neural network $b$-jet tagging algorithm~\cite{ref:nnbtagger}.
The variables used to identify such jets rely on the presence
and characteristics of a secondary vertex and tracks with
high impact parameter inside the jet.

The top quark pair is reconstructed using a 
kinematic fitter~\cite{ref:hitfit}, which
varies the four-momenta of the detected objects
within their resolutions and minimizes a $\chi^2$ statistic,
constraining both \Wdbos\ masses to exactly $80.4\GeV$
and top quark masses to exactly $170\GeV$.
The $b$-tagged jet of highest \pt\  
and the three remaining jets with highest \pt\ are used in the fit.
The $b$-tagging information is used to reduce the
number of jet-parton assignments considered in the fit.
Only events in which the kinematic fit converges are used,
and for each event only the reconstruction 
with the lowest $\chi^2$ is retained. 

The jet-\pt\ selection criteria strongly affect the observed asymmetry
(see Fig.\ \ref{fig:diffAsym}), and this must be 
considered when comparing a model to data.
Fortunately, these effects can be approximated by simple cuts on 
particle-level momenta without changing the asymmetry by more than 2\%
(absolute).
This is verified using several simulated samples with generated asymmetries
and particle jets clustered using the \pxcone\ algorithm~\cite{ref:pxcone}
(``E'' scheme and ${\mathcal R} = 0.5$).
The particle jet cuts are $\pt>21\GeV$ and $|\eta|<2.5$, 
with the additional requirement on the leading particle jet
$\pt>35\GeV$ and the lepton requirements detailed above.
Systematic uncertainties on jet energy calibration
introduce possible shifts of the particle jet thresholds.
The shifts are $^{+1.3}_{-1.5}\GeV$ for the leading jet
and $^{+1.2}_{-1.3}\GeV$ for the other jets, for 
$\pm 1$ standard deviation (sd) changes in the
jet energy calibration.
The resulting changes in the asymmetry predicted using
\mcatnlo\ are of the order of 0.5\%.
The effect of all other selections on the asymmetry is negligible.
The predictions in Table\ \ref{tab:pred} use a more complete description 
of the acceptance based on efficiencies factorized in \pt\ and $\eta$,
accurate to $<1\%$ (absolute).

Misreconstructing  the sign of \dy\ dilutes the asymmetry.
Such dilution can arise from misidentifying lepton charge
or from misreconstructing event geometry.
The rate for misidentification of lepton charge
is taken from the signal simulation and verified using data.
False production asymmetries arising from asymmetries in the rate for
misidentification of lepton charge are negligible owing
to the frequent reversal of the \DZ\ solenoid
and toroid polarities.

\begin{figure}
\vspace{-0.1cm}
	\includegraphics[width=\fullWid\linewidth]{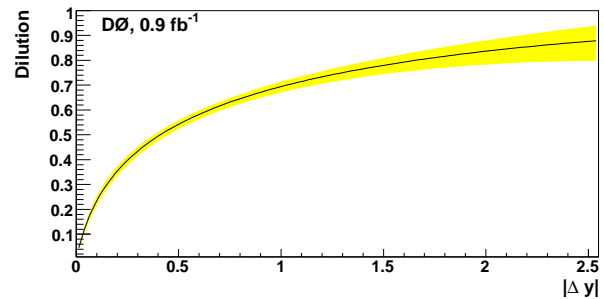}
\vspace{-0.4cm}
\caption{
The geometric dilution and its uncertainty band
as a function of generated \adYgen\ for standard model \ttbarx\ production
and $\gefour$ jets.
}
\label{fig:dil}
\end{figure}

The dilution, \dil, depends mainly on \adYgen. It
is defined as $\dil = 2 P - 1$,
where $P$ is the probability of reconstructing the correct sign of \dy.
It is obtained from
\ttbarx\ events generated with \pythia~\cite{ref:pythia} 
and passed through a \geant-based simulation~\cite{ref:geant} of the
\DZ\ detector, and is parametrized as:
\begin{equation}
\dil\left(\adYgen\right) = c_0 \ln \left(1 + c_1 \adYgen + c_2 \adYgen^2\right),
\label{eq:dil}
\end{equation}
with the parameters given in Table\ \ref{tab:dilpar} (see Fig.\ \ref{fig:dil}).

\begin{table}
\caption{\label{tab:dilpar}Parameters of the dilution.
The $\pm1$ sd values include both statistical and
systematic uncertainties.}
\begin{ruledtabular}
\begin{tabular}{lddd}
Variation & \multicolumn{1}{c}{$c_0$} & \multicolumn{1}{c}{$c_1$} & \multicolumn{1}{c}{$c_2$} \\ 
\hline
$\njet\gefour$    & 0.262 & 14.6 & -1.5 \\
$+1$ sd variation & 0.229 & 20.3 &  1.2  \\
$-1$ sd variation & 0.289 & 11.4 & -2.2 \\
\hline
$\njet=4$         & 0.251 & 17.6 & -1.4  \\  
$+1$ sd variation & 0.201 & 30.3 &  7.7  \\
$-1$ sd variation & 0.293 & 11.6 & -2.3 \\
\hline
$\njet\gef$       & 0.254 &  9.6 &  0  \\ 
$+1$ sd variation & 0.206 & 17.4 &  2.4  \\
$-1$ sd variation & 0.358 &  5.0 & -0.9 \\
\end{tabular}
\end{ruledtabular}
\end{table}

As this measurement is integrated in \adYgen, 
the dependence of the dilution on \adYgen\ introduces
a model dependence into any correction 
from observed asymmetry (\aobs) to a particle-level asymmetry.
Such a correction factor would depend
not only on the model's \adYgen\ distribution, 
but also on its prediction of $\asym\left(\adYgen\right)$. 
Furthermore, such a correction would be sensitive to small
new physics components of the selected sample.
We therefore present a measurement uncorrected for 
reconstruction effects and provide the reader with
a parametrization of \dil\ that describes
these effects, to be applied to any model.

The dilution depends weakly on other variables
correlated with \asym, such as the number of jets.
This possible bias is included in
the systematic uncertainties.
Non-standard production mechanisms can affect reconstruction quality,
primarily due to changes in the momenta of the top quarks.
By studying extreme cases, we find that when comparing
non-standard \ttbarx\ production to data 
an additional 15\% relative uncertainty on 
\asym\ is needed.

The main background is from \wpj\ production.
To estimate it,
we define a likelihood discriminant \ld\ 
using variables that are well-described in our simulation, provide separation
between signal and \wpj\ background, and do not bias \ady\ for the selected signal.
The following variables are used:
the \pt\ of the leading $b$-tagged jet,
the $\chi^2$ statistic from the kinematic fit,
the invariant mass of the jets assigned to the hadronic \Wdbos\ decay,
and $k_{T}^{\min} = p_{T}^{\min} R^{\min}$, where $R^{\min}$ is the smallest angular distance
between any two jets used in the kinematic fit, and $p_{T}^{\min}$ is
the smaller of the corresponding jets' transverse momenta.

The next largest background 
after \wpj\, is from multijet production, 
where a jet mimics an isolated \eom. 
Following the procedure described in Ref.\ \cite{ref:matrixmethod},
the distributions in likelihood discriminant and reconstructed asymmetry 
for this background are derived from samples of data that fail 
lepton identification.
The normalization of this background is estimated from the 
size of those samples and the 
large difference in efficiencies of lepton identification
for true and false leptons.
The effects of additional background sources not considered
explicitly in extracting \asym; namely \zpj, 
single top quark, and diboson production; 
are evaluated using ensembles
of simulated datasets and found negligible.

\begin{figure*}
\begin{center}
\begin{tabular}{l@{\hspace{.4cm}}r}
\includegraphics[viewport=0 0 528 331, clip, width=\halfWid\linewidth]{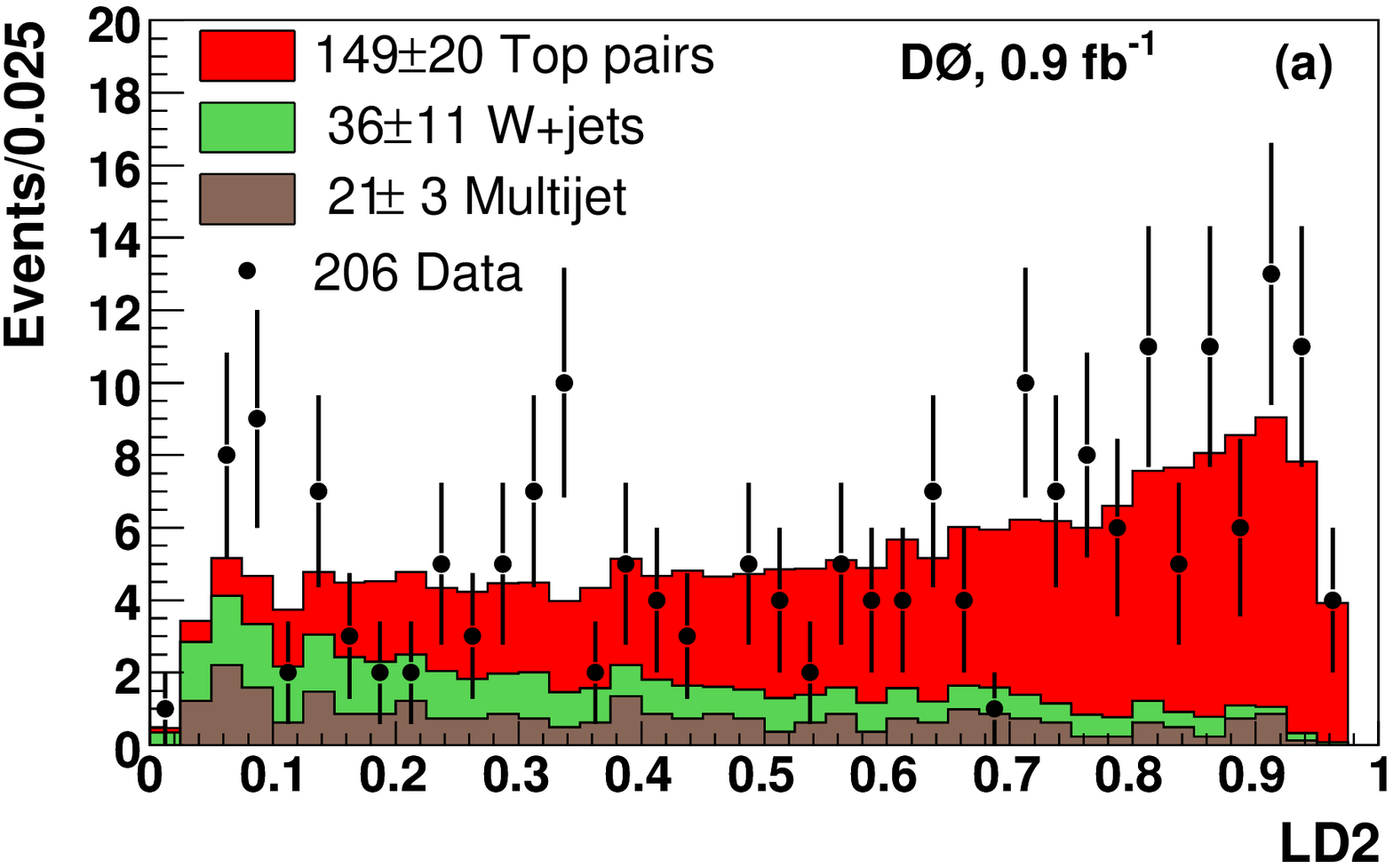} &
\includegraphics[viewport=0 0 528 331, clip, width=\halfWid\linewidth]{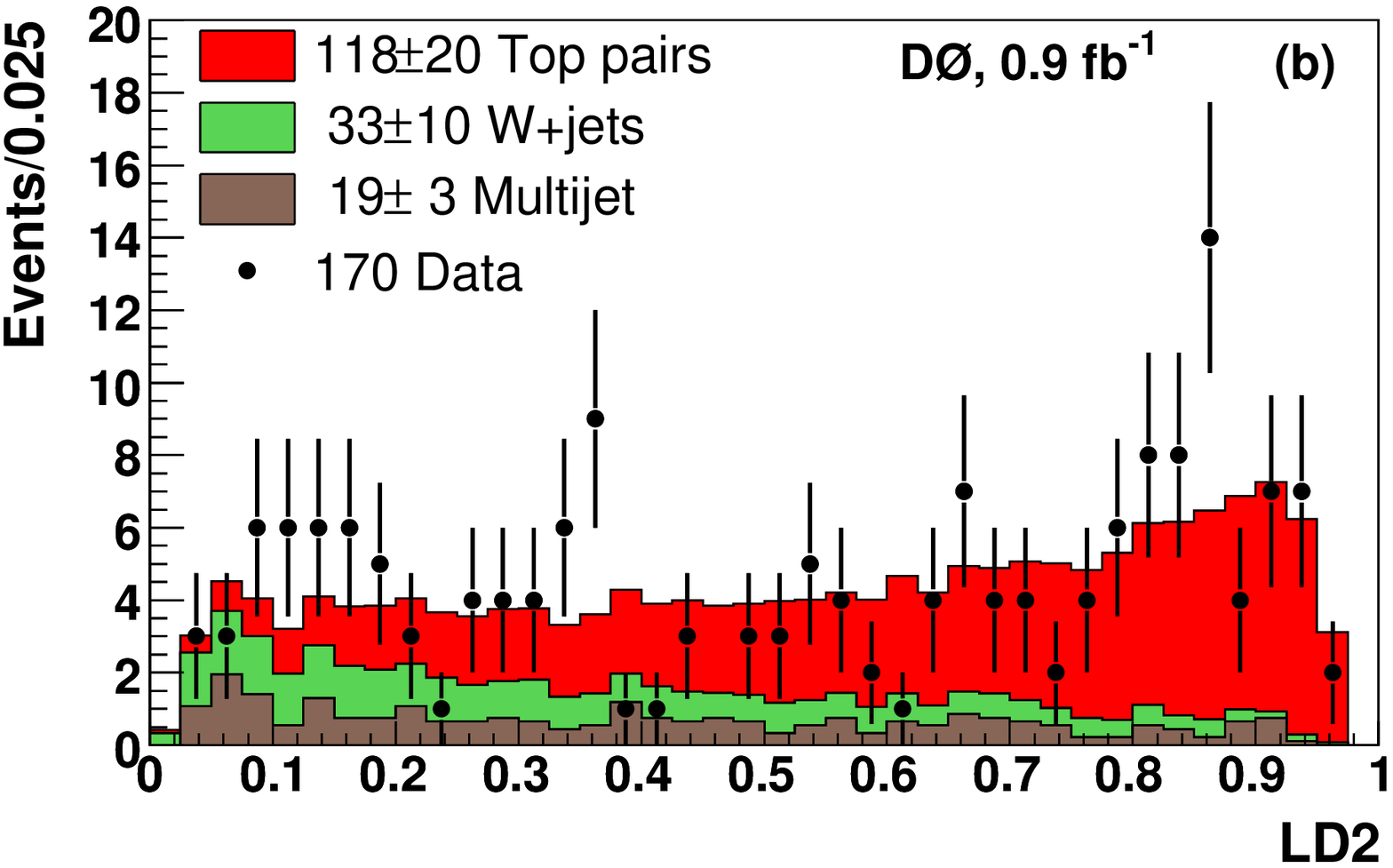} \\
\end{tabular}
\end{center}
\vspace{-0.6cm}
\caption{
Comparison of data for \gefour\ jets with the fitted model
 as a function of \ld\ 
for events reconstructed (a) as forward ($\dyrec>0$)
and (b) as backward ($\dyrec<0$). 
The number of events from each source is listed
with its statistical uncertainty.}
\label{fig:FvsB}
\end{figure*}

The sample composition and \asym\ are extracted from a
simultaneous maximum-likelihood fit to data
of a sum of contributions to \ld\ and
to the sign of the reconstructed \dy\ 
(\dyrec) from
forward signal, backward signal, \wpj, and
multijet production.
Both signal contributions are generated with \pythia,
have the same distribution in \ld,
and differ only in their being reconstructed as either forward or backward.
The \wpj\ contribution is generated with \alpgen~\cite{ref:alpgen} 
interfaced to \pythia\ and has its own reconstructed asymmetry.
Although \Wdbos\ production is inherently asymmetric,
the kinematic reconstruction to the \ttbarx\ hypothesis
reduces its reconstructed asymmetry to $\left[\WArecoSc\stat\right]\%$.
The multijet contribution is derived from data, as described above.
The fitted parameters are shown in Table\ \ref{tab:fitres}.
Correlations between the asymmetry and the other parameters
are $<10\%$. 
The fitted asymmetries in data are consistent with 
the SM predictions given in Table\ \ref{tab:pred}.
In Fig.\ \ref{fig:FvsB} we compare the 
fitted distributions 
to data for events with \gefour\ jets.

\begin{table}
\caption{\label{tab:fitres}Number of selected events and fit results in data.
}
\begin{tabular*}{\linewidth}{l*{3}{@{\extracolsep{\fill}}r@{\extracolsep{0pt}}l}}
\hline
\hline
& \multicolumn{2}{c}{$\gefour$ Jets} & \multicolumn{2}{c}{4 Jets} & \multicolumn{2}{c}{\gef\ Jets}  \\ 
\hline
No. Events  &     \multicolumn{2}{c}{376} & \multicolumn{2}{c}{308} & \multicolumn{2}{c}{68} \\ 
\hline 
\ttbarx   & \multicolumn{1}{r@{}}{$266$} & $^{+23}_{-22}$ &  
	 \multicolumn{1}{r@{}}{$214$} & $\pm20$    & \multicolumn{1}{r@{}}{$54$} & $^{+10}_{-12}$ \\
\wpj    & \multicolumn{1}{r@{}}{$70$}  & $\pm21$  & 
	 \multicolumn{1}{r@{}}{$61$}  & $^{+19}_{-18}$    & \multicolumn{1}{r@{}}{$7$}&$^{+11}_{-5}$ \\
Multijets    & \multicolumn{1}{r@{}}{$40$}  & $\pm4$     &
	\multicolumn{1}{r@{}}{$32.7$} & $^{+3.5}_{-3.3}$ & \multicolumn{1}{r@{}}{$7.1$}&$^{+1.6}_{-1.5}$ \\
\asym  &  \multicolumn{1}{r@{}}{$( \nomA $} & $\fitA  )\%$ &
	  \multicolumn{1}{r@{}}{$( \nomAF$} & $\fitAF )\%$ &
	  \multicolumn{1}{r@{}}{$( \nomAM$} & $\fitAM )\%$ \\ 	
\hline
\hline
\end{tabular*}
\end{table}

The dominant sources of systematic uncertainty for the measured asymmetry
are the relative jet energy calibration between data and simulation ($\pm0.5\%$),
the asymmetry reconstructed in \wpj\ events ($\pm0.4\%$),
and the modeling of additional interactions during a single 
\ppbar\ bunch crossing ($\pm0.4\%$).
The total systematic uncertainty for the asymmetry is $\systA\%$,
which is negligible compared to the statistical uncertainty.

We check the simulation of the production asymmetry, and
of the asymmetry reconstructed under the \ttbarx\ hypothesis
in the \wpj\ background, by repeating the analysis in 
a sample enriched in \wpj\ events. The selection criteria
for this sample are identical to the main analysis, 
except that we veto on any $b$-tags.
Both the fully reconstructed asymmetry and the
forward-backward lepton asymmetry are consistent with 
expectations.
We also find that the fitted sample
composition (Table\ \ref{tab:fitres}) is consistent
with the cross section for \ttbarx\ production obtained
in a dedicated analysis on this dataset.
We check the validity of the fitting procedure, 
its calibration, and its statistical uncertainties using 
ensembles of simulated datasets.

\begin{figure}
\vspace{-0.1cm}
	\includegraphics[width=\fullWid\linewidth]{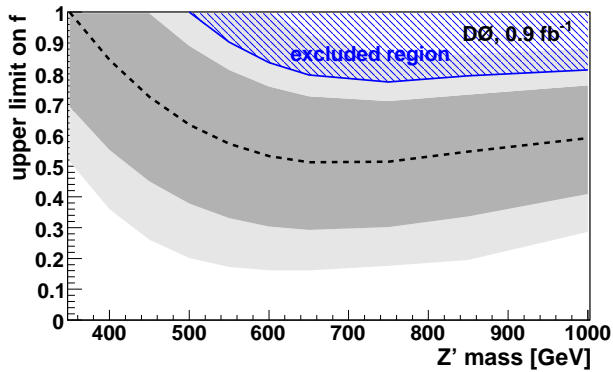}
\vspace{-0.4cm}
\caption{
95\% C.L. limits on the fraction of \ttbar\ produced via a \Zp\ resonance
as a function of the \Zp\ mass, under assumptions detailed in the text.
Limits expected in the absence of a \Zp\ resonance are shown 
by the dashed curve, with the shaded bands 
showing limits one and two standard deviations away.
The observed limits are shown by the solid curve, and the excluded region 
is hatched.
}
\label{fig:zlim}
\end{figure}

To demonstrate the measurement's sensitivity to new physics,
we examine \ttbar\ production via
neutral gauge bosons (\Zp)
that are heavy enough to decay to on-shell top and antitop quarks.
Direct searches have placed limits on
\ttbar\ production via a heavy narrow resonance~\cite{ref:resonance},
while the asymmetry in \ttbar\ production may be
sensitive to production via both narrow and wide resonances.
The $\Zp\to\ttbar$ channel is of interest in models with a ``leptophobic''
\Zp\ that decays dominantly to quarks. 
We study the scenario where the coupling between the \Zp~boson and quarks is
proportional to that between the \Z~boson and quarks, and interference
effects with SM \ttbar\ production are negligible.
Using \pythia\ we simulate \ttbar\ production 
via \Zp\ resonances with  decay rates chosen to yield 
narrow resonances as in Ref.\ \cite{ref:resonance},
and find
large positive asymmetries $[(13$--$35)\%]$,
which are a consequence of the predominantly left-handed decays.
We predict the distribution of \asym\
as a function of 
the fraction ($f$) of \ttbar\ events produced via a \Zp\ resonance
of a particular mass from ensembles of simulated datasets.
We use the procedure of Ref.\ \cite{ref:FC}
to arrive at the limits shown in Fig.\ \ref{fig:zlim}.
These limits can be applied to wide \Zp\ resonances by averaging
over the distribution of \Zp\ mass.

In summary, we present the first measurement of the integrated
forward-backward charge asymmetry in \ttbarx\ production.
We find that acceptance affects the asymmetry and must
be specified as above, 
and that corrections for reconstruction effects are 
too model-dependent to be of use.
We observe an uncorrected asymmetry of
$\aobs = \left[\nomA \fitA \stat \systA \syst \right]\%$
for \ttbarx\ events with \gefour\ jets that are within
our acceptance, and we provide a dilution function (Eq.\ \ref{eq:dil})
that can be applied to any model (through Eq.\ \ref{eq:applyDil}).
For events with only four jets and for those
with \gef\ jets, we find
$\aobs =  \left[ \nomAF \fitAF \stat \systAF \syst \right]\% $ and 
$\aobs =  \left[ \nomAM \fitAM \stat \systAM \syst \right]\%$, 
respectively,
where most of the systematic uncertainty is from 
migrations of events between the two subsamples.
The measured asymmetries are consistent with the
\mcatnlo\ predictions for standard model production.

%
We thank the staffs at Fermilab and collaborating institutions, 
and acknowledge support from the 
DOE and NSF (USA);
CEA and CNRS/IN2P3 (France);
FASI, Rosatom and RFBR (Russia);
CAPES, CNPq, FAPERJ, FAPESP and FUNDUNESP (Brazil);
DAE and DST (India);
Colciencias (Colombia);
CONACyT (Mexico);
KRF and KOSEF (Korea);
CONICET and UBACyT (Argentina);
FOM (The Netherlands);
Science and Technology Facilities Council (United Kingdom);
MSMT and GACR (Czech Republic);
CRC Program, CFI, NSERC and WestGrid Project (Canada);
BMBF and DFG (Germany);
SFI (Ireland);
The Swedish Research Council (Sweden);
CAS and CNSF (China);
Alexander von Humboldt Foundation;
and the Marie Curie Program.
%


\begin{thebibliography}{99}
%
\bibitem[a]{alton}
Visitor from Augustana College, Sioux Falls, SD, USA.
\bibitem[b]{burdin}
Visitor from The University of Liverpool, Liverpool, UK.
\bibitem[c]{podesta-lerma}
Visitor from ICN-UNAM, Mexico City, Mexico.
\bibitem[d]{quadt,meyer}
Visitor from II. Physikalisches Institut, Georg-August-University, G{\"o}ttingen, Germany.
\bibitem[e]{voutilainen}
Visitor from Helsinki Institute of Physics, Helsinki, Finland.
\bibitem[f]{wenger}
Visitor from Universit{\"a}t Z{\"u}rich, Z{\"u}rich, Switzerland.

\bibitem[\dag]{IntFellows}
Fermilab International Fellow.
\bibitem[\ddag]{deceased}
Deceased.

%
\vskip 0.25cm

\bibitem{ref:asymwhy}
J.\,H. K\"uhn and G. Rodrigo, 
Phys.\ Rev.\  D {\bf 59}, 054017 (1999).

\bibitem{ref:asymtev}
M.\,T. Bowen \etal, Phys. Rev. D {\bf 73}, 14008 (2006).

\bibitem{ref:asymNNLO}
S. Dittmaier \etal, Phys. Rev. Lett. {\bf 98}, 262002 (2007). 

\bibitem{ref:axigluon}
For example, O. Antu\~{n}ano, J.\,H. K\"uhn and G. Rodrigo,
arXiv:0709.1652.

\bibitem{ref:eta} Rapidity $y$ and pseudorapidity $\eta$ are defined as
functions of the polar angle $\theta$ as
$y(\theta,\beta) \equiv
{1 \over 2} \ln{[(1+\beta\cos{\theta})/(1-\beta\cos{\theta})]};
\eta(\theta) \equiv y(\theta,1)$,
where $\beta$ is the ratio of a particle's momentum to its energy. 

\bibitem{ref:d0det}
V.\,M. Abazov \etal\ (D0 Collaboration), 
Nucl. Instrum. Methods A {\bf 565}, 463 (2006).

\bibitem{ref:mcatnlo}
S. Frixione and B.R.\,Webber, JHEP {\bf 0206}, 29 (2002);\\
S. Frixione \etal, JHEP {\bf 0308}, 7 (2003).

 \bibitem{ref:d0jets}
G.\,C. Blazey \etal, in
{\sl Proceedings of the Workshop: QCD and Weak Boson
Physics in Run II,} edited by U.~Baur, R.K.~Ellis, and
D. Zeppenfeld, Fermilab-Pub-00/297 (2000).

\bibitem{ref:matrixmethod}
V.\,M. Abazov \etal\ (\DZ\ Collaboration), Phys. Rev. D {\bf 76}, 092007 (2007).

\bibitem{ref:nnbtagger}
T. Scanlon, Ph.D. thesis, University of London (2006), Fermilab-Thesis-2006-43.

\bibitem{ref:hitfit}
S. Snyder, Ph.D. Thesis, State University of New York at Stony Brook (1995).

\bibitem{ref:pxcone}
C. Adloff \etal\ (H1 Collaboration), Nucl. Phys. {\bf B545}, 3 (1999).

\bibitem{ref:pythia}
T. Sj\"{o}strand \etal, Comput. Phys. Commun. {\bf 135}, 238 (2001).

\bibitem{ref:geant} 
R. Brun and F. Carminati, CERN Program Library Long Writeup W5013, 1993 (unpublished).

\bibitem{ref:alpgen} 
M.\,L. Mangano \etal, JHEP {\bf 0307}, 1 (2003); \\
S. H\"{o}che \etal, arXiv:hep-ph/0602031.

\bibitem{ref:PDG}
W.-M. Yao \etal, J.\ Physics G {\bf 33}, 1 (2006).

\bibitem{ref:resonance}
V.\,M. Abazov \etal\ (\DZ\ Collaboration), Phys. Rev. Lett. {\bf 92}, 221801 (2004);
T. Aaltonen \etal\ (\CDF\ Collaboration), arXiv:0709.0705.

\bibitem{ref:FC}
G. Feldman and R. Cousins, Phys. Rev. D {\bf 57}, 3873 (1998).

\end{thebibliography}
\end{document}